# Test experiments with distributed acoustic sensing and hydrophone arrays for locating underwater sound sources.

Jörg Rychen[1*], Patrick Paitz[2], Pascal Edme[2], Krystyna Smolinski[2], Joeri Brackenhoff[2], and Andreas Fichtner[2*]

[1] Department of Information Technology and Electrical Engineering, ETH Zurich, Zurich, Switzerland
[2] Department of Earth Sciences, ETH Zurich, Zurich, Switzerland
* Corresponding authors: jrychen@ethz.ch, andreas.fichtner@erdw.ethz.ch

Date: June 2023

Dataset:
https://zenodo.org/7886409

Git repository:
https://gitlab.switch.ch/cetacean-communication/expeditions/daslakezh/datapublication.git

## Abstract

Whales and dolphins rely on sound for navigation and communication, making them an intriguing subject for studying language evolution. Traditional hydrophone arrays have been used to record their acoustic behavior, but optical fibers have emerged as a promising alternative. This study explores the use of distributed acoustic sensing (DAS), a technique that detects local stress in optical fibers, for underwater sound recording. An experiment was conducted in Lake Zurich, where a fiber-optic cable and a self-made hydrophone array were deployed. A test signal was broadcasted at various locations, and the resulting data was synchronized and consolidated into files. Analysis revealed distinct frequency responses in the DAS channels and provided insights into sound propagation in the lake. Challenges related to cable sensitivity, sample rate, and broadcast fidelity were identified. This dataset serves as a valuable resource for advancing acoustic sensing techniques in underwater environments, especially for studying marine mammal vocal behavior.

## Background

Whales and dolphins rely on sound for navigation and communication in their underwater world, and their social and cognitive abilities make them a compelling subject in evolutionary biology to research the evolution of language. To record the acoustic behavior of these marine mammals, hydrophone arrays have been used to locate and track animals in the wild (Barlow et al., 2018; Miller & Dawson, 2009; Møhl et al., 2001). However, recent developments have demonstrated that optical fibers can serve as a viable alternative to hydrophones for this purpose (Bouffaut et al., 2022).

Distributed acoustic sensing (DAS) is a technique that utilizes an optical fiber probed with a laser interrogator to detect local stress along the fiber, providing high channel-count acoustic measurements over long distances. Compared to hydrophone arrays, DAS seams easier to deploy underwater and offers better beamforming with more channels. However, several questions arise regarding the sensitivity of the fiber, its noise floor, or the maximum achievable sample rate. Does the increased channel count compensate for the increased noise level compared to hydrophones? How accurate can sources be localized using DAS in practice? Can the fiber itself be localized using controlled sources?

To address these questions, we conducted an experiment and acquired a test dataset using DAS in the Lake Zurich with controlled broadcasts of a test signal. In this unique test situation, we also deployed our own self-made hydrophone array to gather extensive calibration and ground truth data. This report provides the description of the instrumentation, experimental procedures, data consolidation process, resulting file format, and an overview of the data with first insights.

## Experiment

For our experiment, we installed a 1000 m long fiber-optic cable along the bottom of Lake Zurich at depth of 2 - 18 m. The cable was laid out in a looped formation, crossing itself at a right angle (Figure 1). A DAS interrogator sampled the strain in the fiber with a channel spacing of 1 m and a sample rate of 5 kHz. See paragraph *Test site* and *Fiber layout.* Field work impressions can be found at [https://youtu.be/nU1Gro0DH_4](https://youtu.be/nU1Gro0DH_4)

During the experiments, a hydrophone array was deployed in the lake that consisted of three drifting buoys, each with a tetrahedral four-hydrophone array hanging 6 - 12 m below the buoy. See paragraph *Hydrophone array recording system.*

We designed a 20 s long test signal that had 3 s of white noise, 10 s of linear frequency sweep from 100 Hz to 4 kHz, 5 pure tones of 1 s in the same frequency range, and two sharp pulses at the beginning and end of the signal. See the top two graphs of Figure 2 and paragraph *Test signal.*

For the experiment we broadcasted the test signal in the lake using an underwater load speaker deployed from a rowboat. See paragraph *Underwater speaker system*. A series of broadcasts at various locations around the test installation forms an experimental dataset. See paragraph *Experimental runs.*

The broadcast system and hydrophone recorders yielded data synchronized to the GPS clock with sub-sample precision. The DAS interrogator yielded also data synchronized with a GPS timer.

After the experiments we consolidated the data into files of 25 s duration for each broadcast (see paragraph *Data management)*. Each file contains the signal of the speaker (i.e. the test signal), the hydrophone signals (3x4 channels, 50 kHz sample rate), and the DAS data (946 channels, 5 kHz sample rate). Additionally, each file contains the positions of the source and hydrophone recorders and further relevant metadata (see paragraph *HDF5 file format*). The published dataset comprises 93 files in 5 series, totaling 25 GB in size and is available on Zenodo (Rychen et al., 2023).

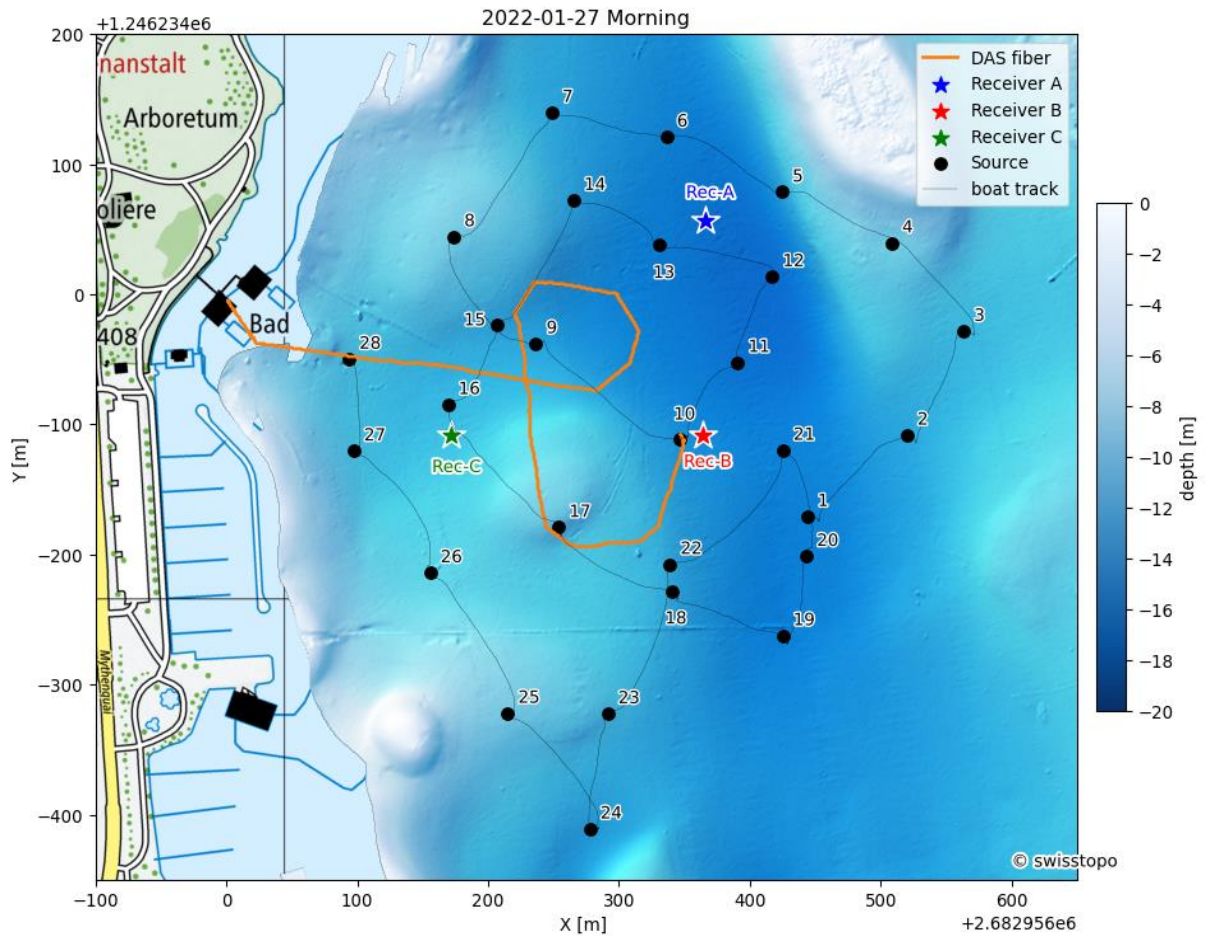

*Figure 1. Overview of the experimental setup and the locations of the test signal broadcasting. The orange line represents the layout of the fiber-optic cable on the bottom of Lake Zurich. The three stars indicate the positions of the floating hydrophone recorders. The black dots with serial numbers represent the locations where the test signal was broadcasted. The thin black line shows the path taken by the rowboat. "Bad" (Seebad Enge) is the good place where the interrogator was installed. Shown is the dataset from 27th Jan. 2022 comprising 28 broadcasts.*

## Results

Figure 2 provides an overview of the signals in one file, including spectrograms of the source signal, hydrophone signal, and DAS channels. We observed a clear response to the broadcast sound in DAS channels 0-150 and 800-945, while channels 150-800 showed almost no response. The sensitive channels show each a distinct frequency response that is almost independent of the direction of the source, as depicted in Figure 3.

Synchronized recordings of the hydrophones enabled us to analyze the sound wave propagation in the shallow lake. Figure 4 displays the arrival of the first synchronization pulse at the three hydrophone recorders. The impulse response functions from the source to the hydrophones was computed with the white noise sections of the test signal and show the multipath arrival due to reflections on the surface and bottom.

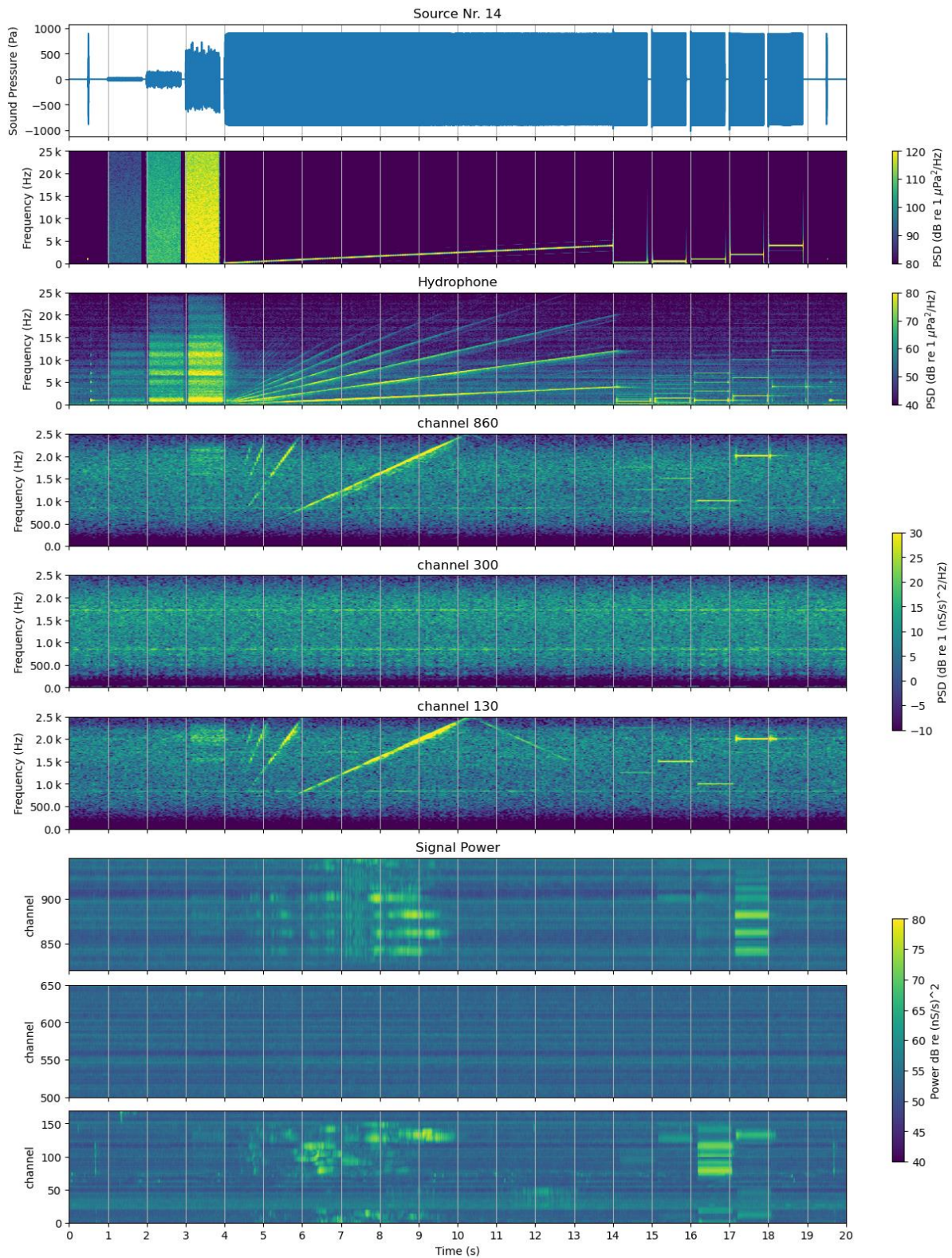

*Figure 2. Signal overview for shot number 14. The top plot is the waveform of the source signal, below the spectrogram of the source signal that shows the white noise sections, the frequency sweep, and the constant frequency sections. The third plot is a spectrogram of one hydrophone signal from recorder A. Two important effects are visible: First, the speaker system exhibits harmonic distortion (non-linearity) noticeable by the overtones of the frequency sweep. Second, the noise sections show a frequency dependence due to the frequency variable gain of the speaker and due to interference of reflections. The fourth to sixth plot are spectrograms of single DAS channels. The frequency sweep is clearly visible for some of the DAS channels. Aliasing is observed by the mirror at the Nyquist frequency for the up sweep. The last three plots visualize the power of the DAS channel signals. The channels 150 – 800 do not show any response.*

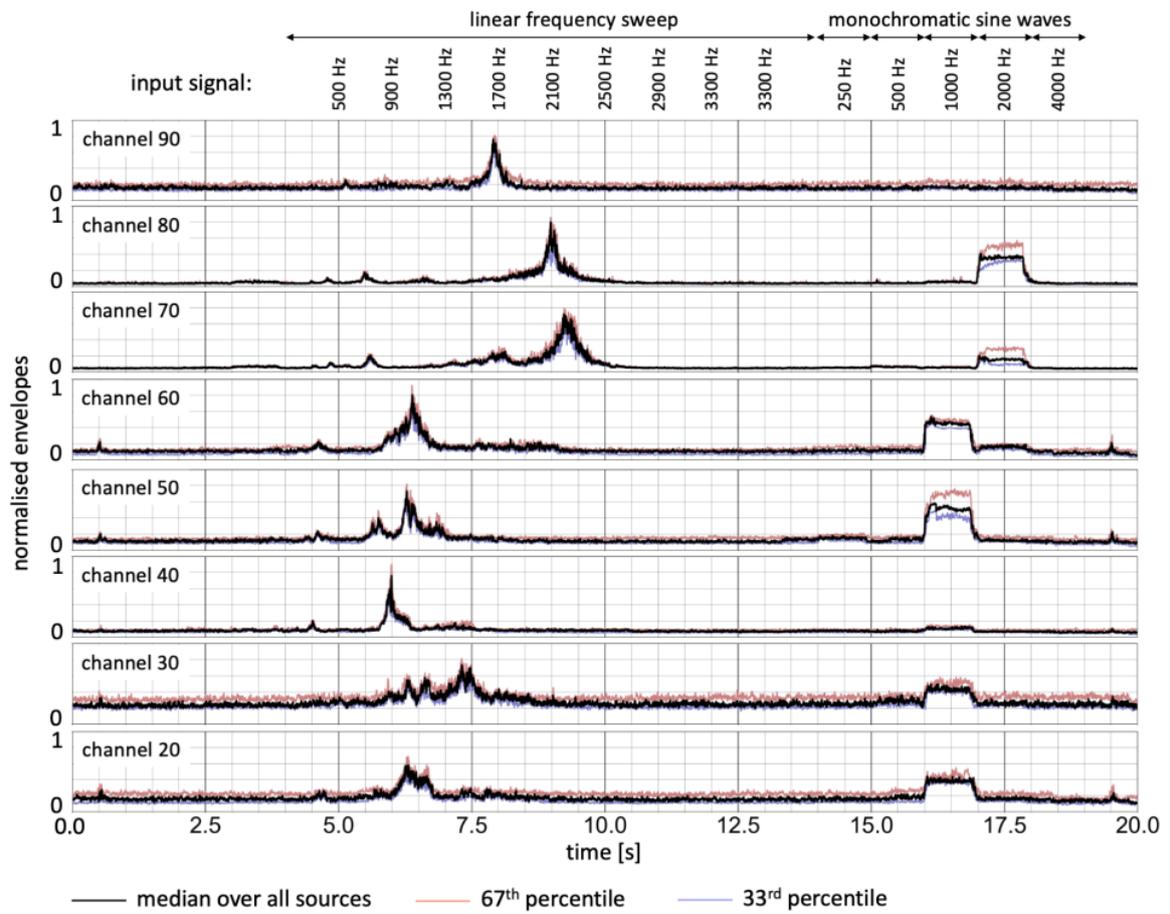

Figure 3. Normalized envelopes of DAS recordings on selected channels for all sources in the experiment, shown in Figure 1. The median is shown in black, the 67th percentile in red, and the 33rd percentile in blue. The pronounced bandpass characteristics are largely independent of the source location and therefore mostly a local property of the medium or fiber.

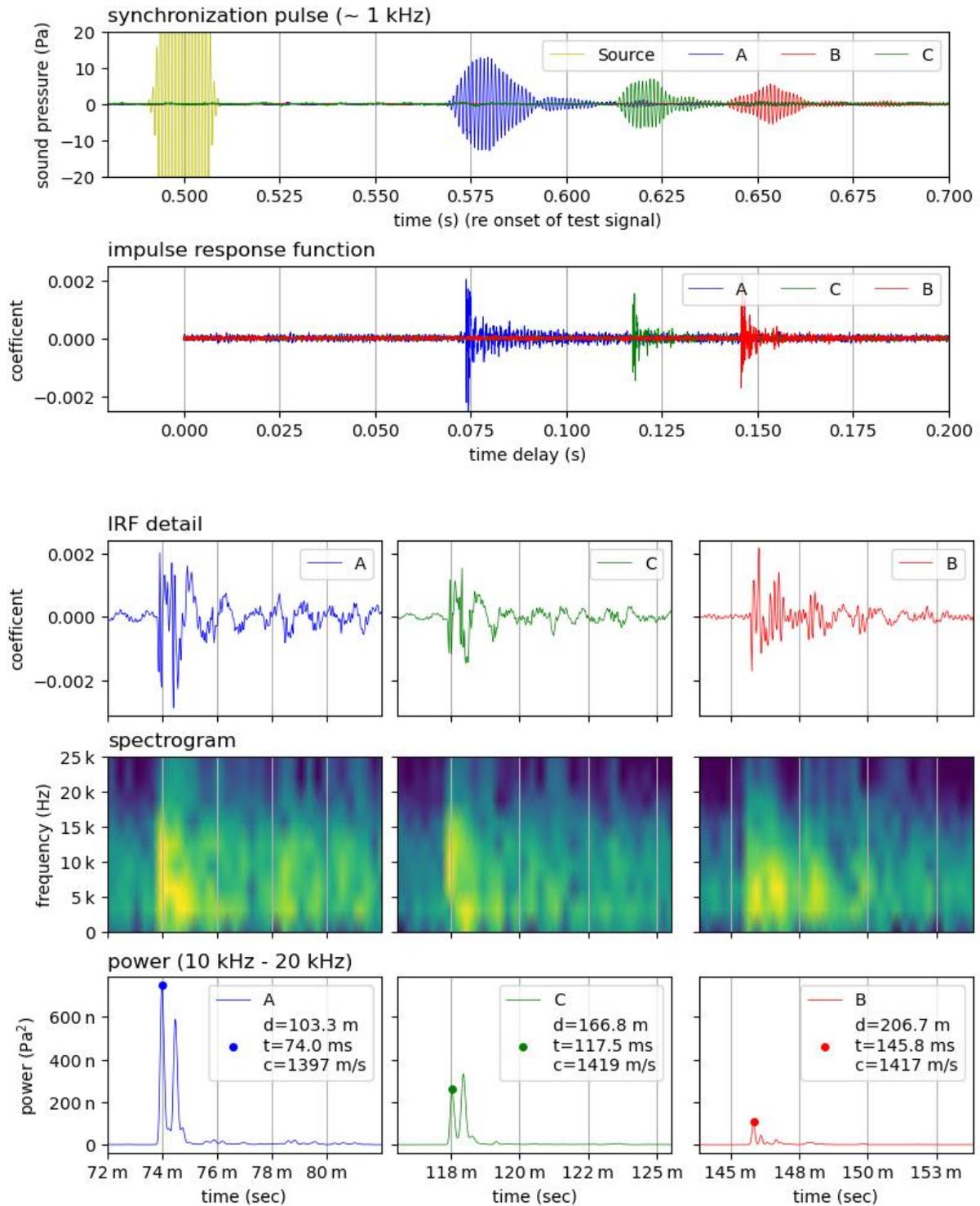

*Figure 4. Hydrophone signals for shot nr. 14. First row: The initial synchronization pulse. Yellow is the source signal and blue, green, red are the arrivals at the hydrophones A, B, C. Second and third row: Measured impulse response functions obtained by cross-correlation of the 1 s long white noise segment of the source signal with the hydrophone signals. Second last row: Spectrographic representation of the impulse response. Last row: Multiple arrivals are more clearly observed when the impulse response is high-pass filtered at 10 kHz. This is to avoid the resonances of the speaker system below 5 kHz (see Figure 11).*

# Discussion

The acquired data allows to asses DAS for recording the vocal behavior of marine mammals. The high channel count of DAS are expected to be an advantage for beamforming to locate the animals and separate their sounds. However, some experimental issues could be identified:

The fiber-optic cable turned out to be insensitive when it sunk into the mud on the ground of the lake. At some locations of the cable we observed very localized (within a few meters) resonances that are independent of the direction of the incident wave. We speculated this could be due to mechanical resonances of the cable or by local resonances due to wave coupling in the ground. The tough cable we used contained the fiber in a gel filled steel tube, but for acoustic sensing in water this might not be the optimal solution in terms of sensitivity.

The experiment highlights the need for a higher sample rate of DAS. The sample rate of 5 kHz in this experiment yields a frequency range up to 2.5 kHz. To record killer whales, a sampling frequency of at least 20 – 50 kHz would be needed.

For beamforming applications it is necessary to know the exact location of the fiber. Controlled broadcasts of source signals, as in this experiment, allow us to compute the location of the fiber with high accuracy.

The experiments have shown the difficulty to broadcast sounds in a broad bandwidth (200 Hz – 20 kHz) with high fidelity (linearity). A better amplifier and a pre whitening filter would improve such experiments by providing a more controlled source signal.

The experimental data is also a ground truth dataset for the hydrophone array and allows for calibrating the tetrahedral arrays by measuring the direction dependent array impulse response functions.

# Methods

## Test site

We conducted our experiment in late January 2022 at the northern tip of Lake Zurich, Switzerland. Covering an area of 90.1 km$^2$, the lake reaches a maximum depth of 136 m and contains 3.9 km$^3$ of fresh water. The lake level was 405.7 m above mean sea level at the time of the experiments. The body of water had a nearly homogeneous temperature of 4.6 °C. The speed of sound is thus 1425 m/s (Grosso & Mader, 2005). Thanks to the very quiet weather conditions, atmospheric influences on the data can also largely be excluded (Figure 5).

We use the official Swiss coordinate system LV95 (EPSG 2056) throughout the dataset (Swisstopo, 2016). The bathymetry dataset is from the Swiss federal office of topography (Swisstopo, 2022). The bathymetry dataset has a 1 m raster and a 1 mm depth resolution (Figure 6).

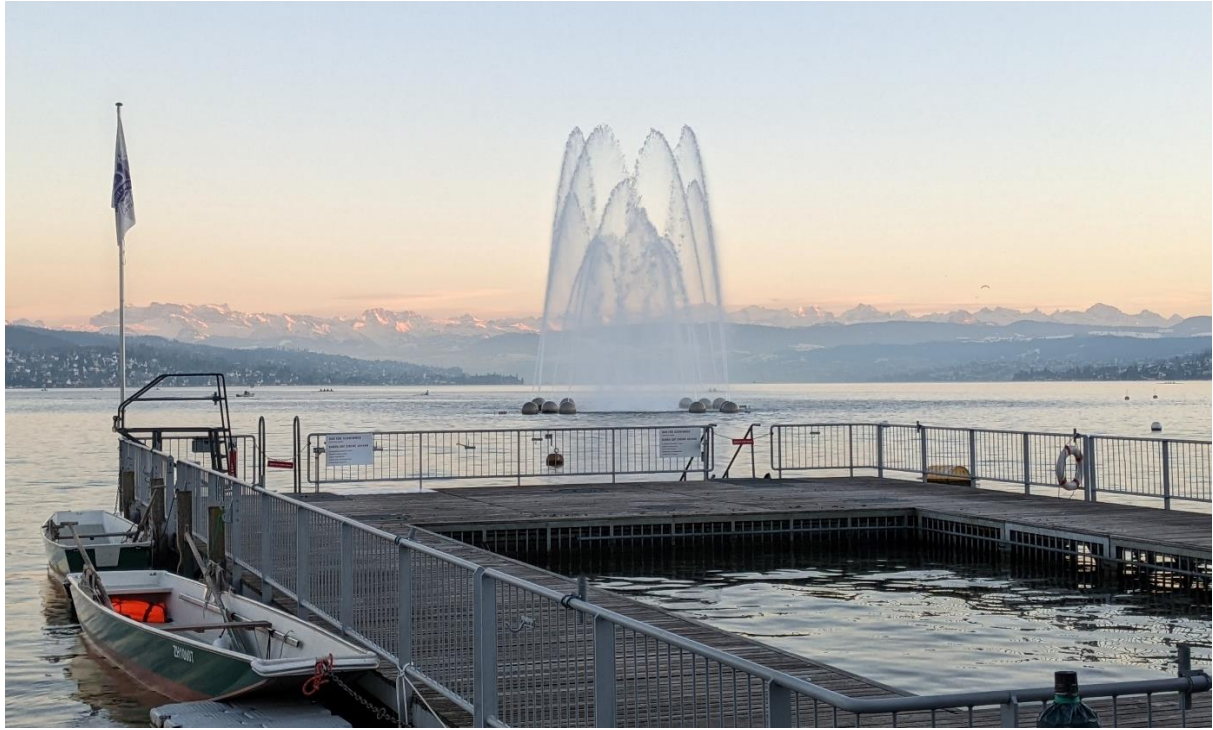

*Figure 5. Lake Zurich photographed from the public bath 'Seebad Enge' on the evening of January 26th 2022. The weather was stable with almost no wind indicated by the flag.*

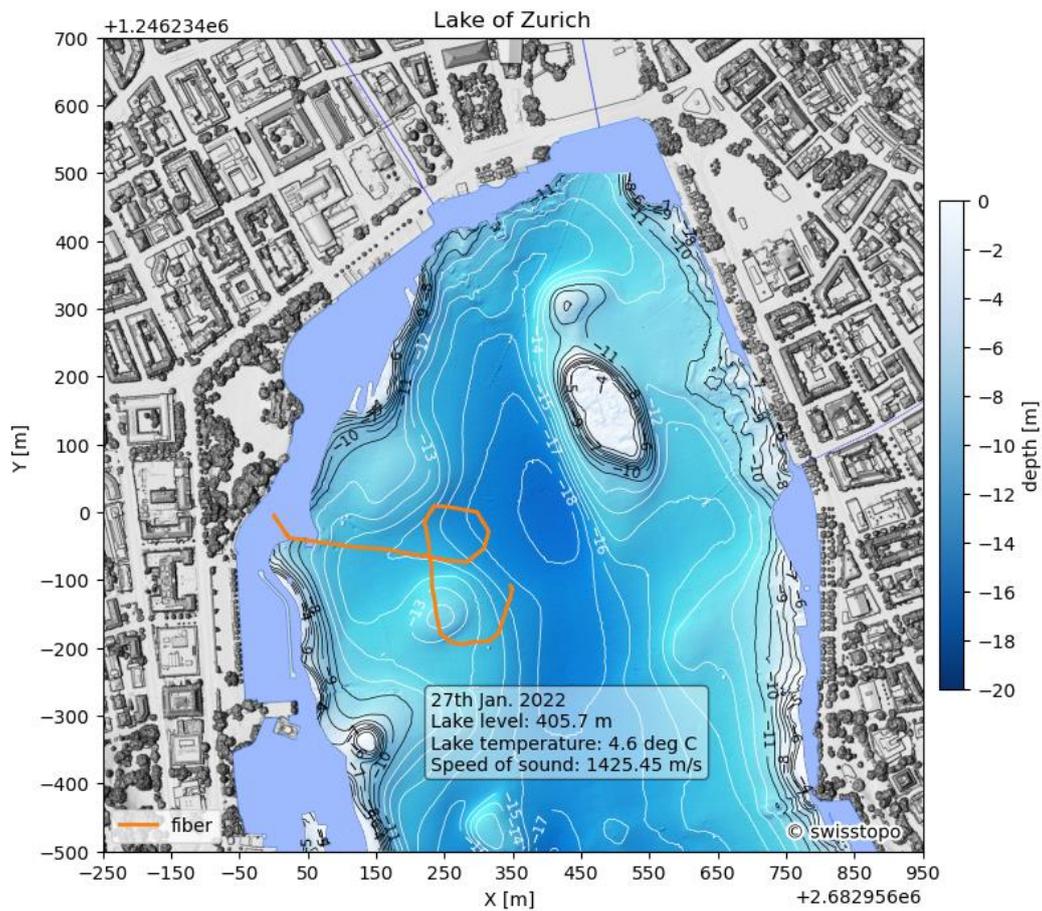

*Figure 6. Overview of the northern tip of Lake Zurich. Inset reports the environmental measurements. Maps and bathymetry data are from* Swisstopo *(Swisstopo, 2022). All coordinates in the official Swiss LV95 (EPSG 2056) coordinate system.*

## Fiber layout

With the assistance of the Zurich Lake Police, we installed 1000 m of Solifos BSAC 4FSC AC3 cable with four single-mode fibers. One of the fibers was connected to a Silixa iDAS v2, which was safely located in the office of the public bath "Seebad Enge." An illustration of the experimental setup is provided in Figure 7. Although our initial plan was to maintain a well-defined geometry, it was challenging to navigate with high accuracy while deploying the cable, resulting in a more irregular shape. During deployment, unavoidable dragging of the cable caused additional distortions.

We estimated the actual DAS channel locations by under-water tap testing during a pleasant dive in the 4°C warm lake water. Along the cable, the water depth varies between 2 m and 18 m, with an average of around 12 m. Throughout its entire length, the cable rested on top or slightly sank into the very soft mud of the lake bottom.

We built special structures into the fiber to experiment with the specialties of DAS: a sphere, a suspension and a bundle, all illustrated in Figure 8.

The DAS channel spacing was 1 m, the gauge length 10 m, and the sampling frequency 5 kHz, leading to around 50 MB of strain rate data per second. We used 946 channels along the 946 m long fiber in the water (Figure 7).

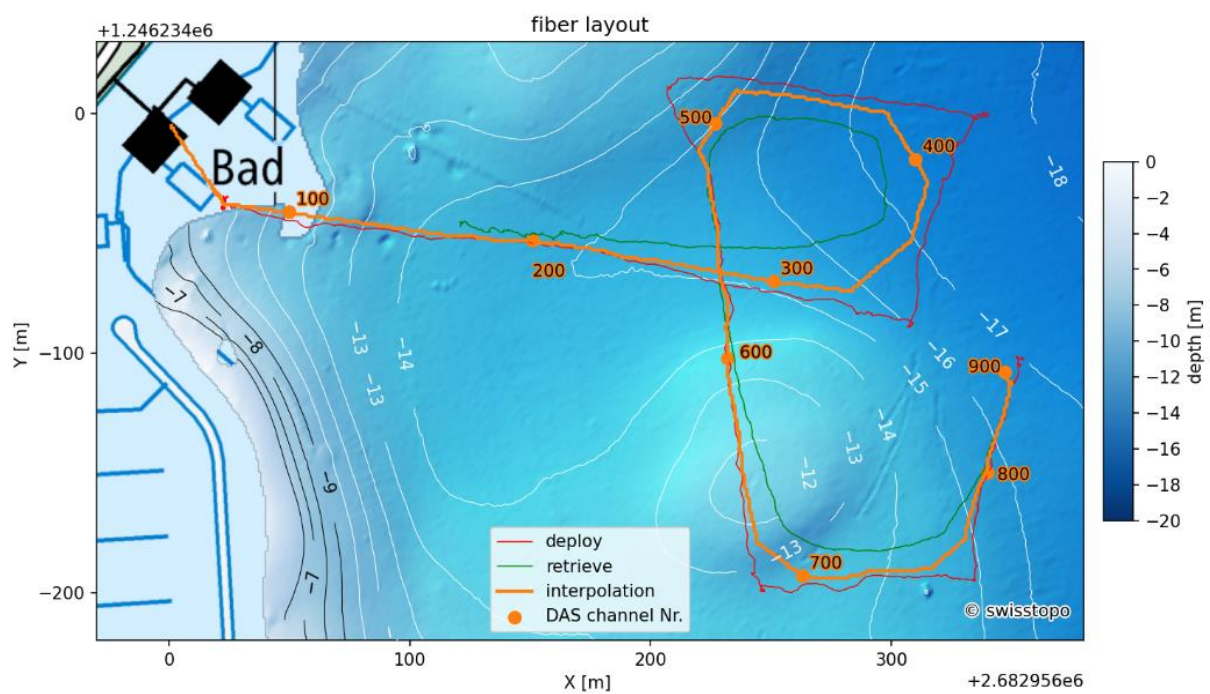

*Figure 7. Layout of the fiber-optic cable on the lake bottom. The boat tracks for deployment and recovery are plotted in red and green. Orange is the interpolated estimate of the fiber layout. The DAS channel numbers are indicated with orange numbers.*

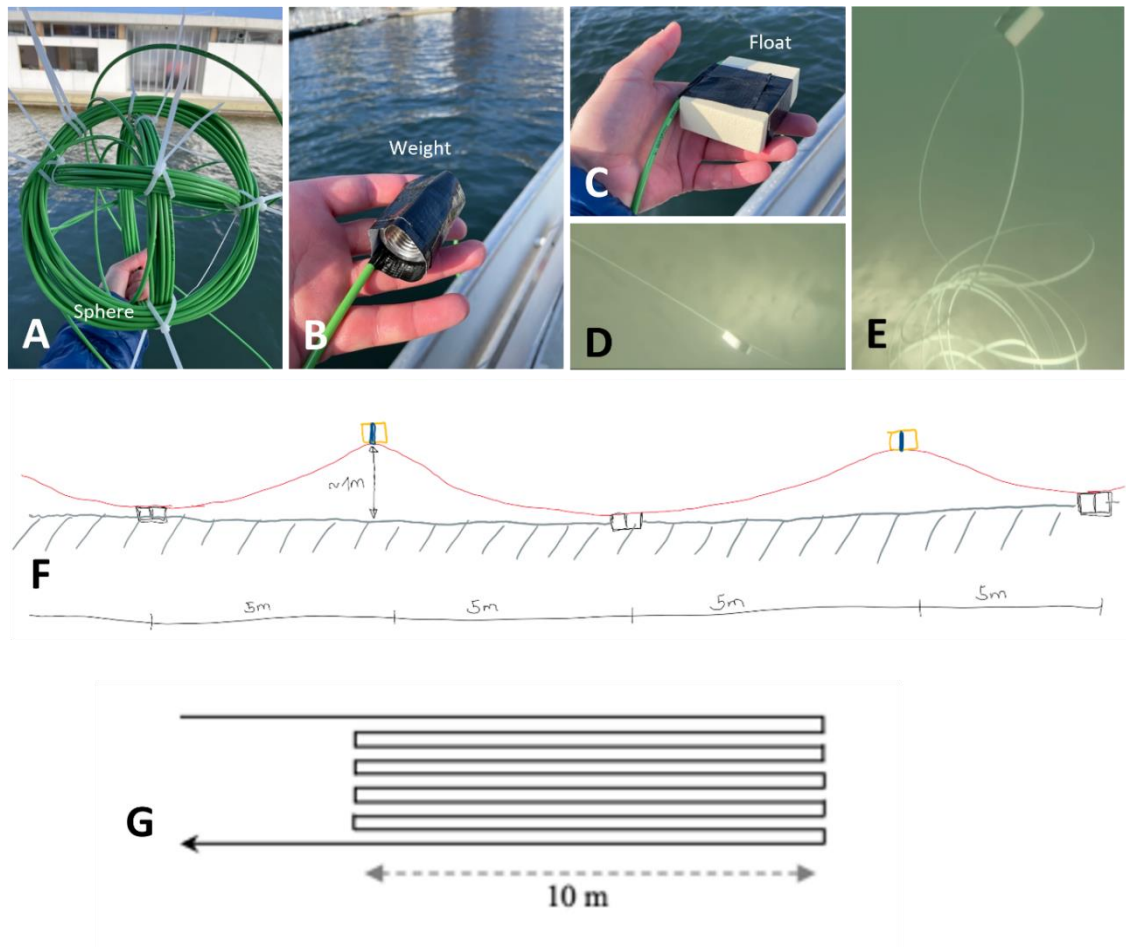

*Figure 8. Special structures in the fiber-optic cable. A) The sphere consisted of three orthogonal coils each with 10 m of cable. B-E) A 50 m long section with alternating weights and floats every 5 m. F) Drawing of the cable suspended above the soft mud on the lake bottom. G) The last 100 m of the cable were looped 10 fold and tied together.*

## Hydrophone array recording system

We used three custom built hydrophone recorders to monitor the propagation of the acoustic waves in the lake. The recorders were mounted in watertight enclosures that acted as free floating buoys (Figure 9). We deployed these buoys around the location of the fiber, with a distance of about 200 m. (Figure 1, labelled A,B and C, ). Each buoy has a GPS receiver to precisely record the position and to synchronize the signal acquisition to a precision of ±100 ns. Each buoy recorded from four hydrophones (HTI-96, High Tec Inc., USA) arranged in a tetrahedron structure of 0.7 m side length hanging 5-8 m below each buoy (Figure 9).

All hydrophones were factory calibrated with sensitivities of -170 dB re 1V/µV (± 0.2 dB) in a frequency range of 2 Hz – 30 kHz. The signals were digitized by 24 bit sigma-delta analog-to-digital converters (NI-9234, National Instruments) at a sampling rate of 51.2 kHz. We resampled all signals at even multiples of 20 µs of the GPS clock, i.e., a sampling rate of 50 kHz. This was implemented on the field programmable gate array (FPGA) of the controller, where we replicated the GPS clock with a phase locked loop (Timekeeper, National Instruments, USA). We then timestamped every sample of the free running AD converter and used linear interpolation for the resampling.

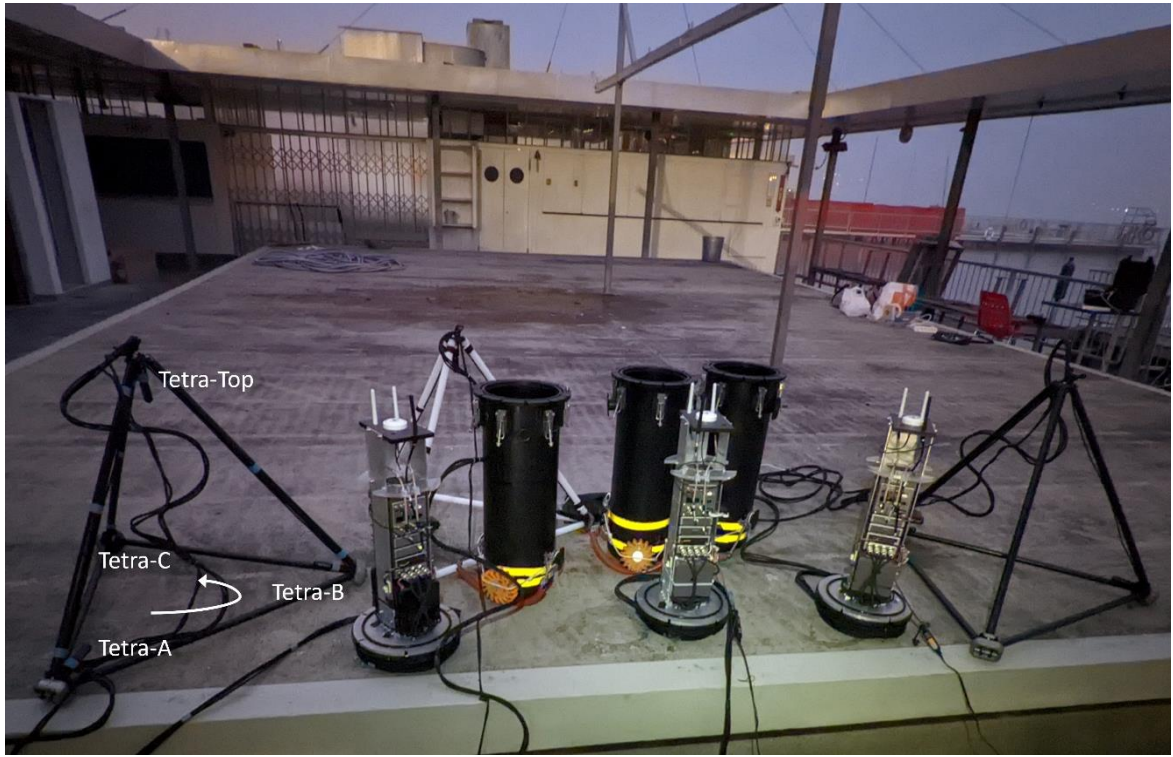

*Figure 9. The three hydrophone recorder buoys. The buoys are opened for maintenance. Each buoy records from an array with four hydrophones arranged in a tetrahedron. The channel names of the hydrophones are indicated. The tetrahedrons are suspended 5 - 15 m below the buoys including a 4 m long elastic rope to dampen the movements from the surface.*

## Underwater speaker system

Our source for acoustic signals was an underwater speaker (Lubell 916, Lubell Labs, USA) that we deployed from a rowboat at various positions and depths (Figure 10). The source was driven by a digital signal controller (CompactRIO, National Instruments, USA) and custom-built amplifier. The source was powered by a 100 Wh lead-acid battery contained in a waterproof transport casing. The source level was up to 180 dB re 1µPa @ 1m and the frequency range 200 Hz to 20 kHz.

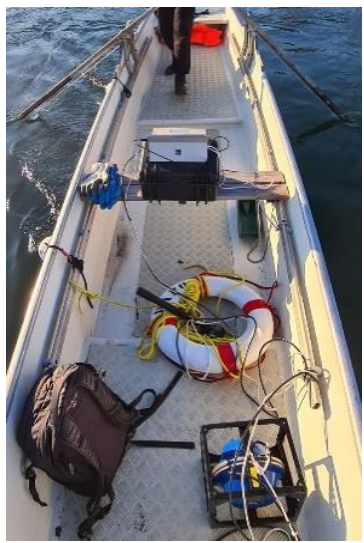

*Figure 10. Installation of the acoustic source on a rowboat. The blue device on the bottom is the underwater speaker. The black box on the bench contains the electronics and batteries. On top of the box is a tablet computer to control the broadcasts.*

In a former expedition we have calibrated the speaker system by recording its output at a distance of 1 m with a calibrated hydrophone. See the provided Jupyter notebook *SpeakerCalibration* for details. We use a constant conversion factor of 180 dB (re 1 µPa / V) which corresponds to 1 kPa / V.

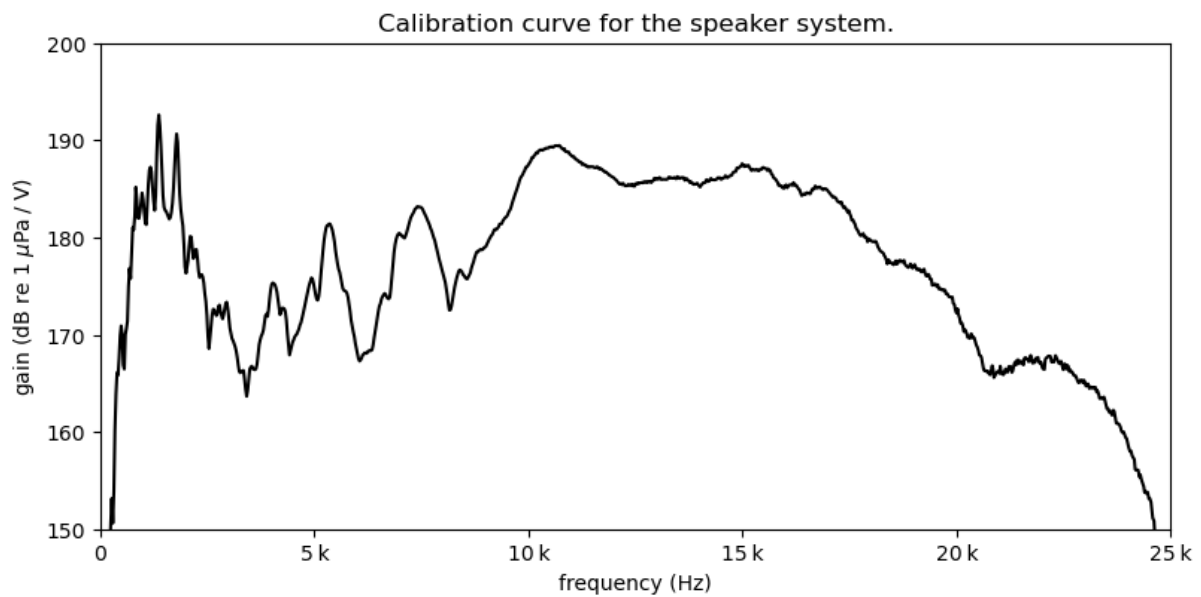

*Figure 11. The gain of the speaker system as a function of frequency. In the range below 4 kHz, we observe two resonances and a drop of gain below 1 kHz. In the range between 10 kHz and 15 kHz, the gain is stable.*

## Test signal

The test signal is a 20 s long audio file with 51.2 kS/s sampling rate and 32 bit values in the range of +- 1. Following an initial synchronization pulse at 1 kHz central frequency, it contains intervals of white noise with increasing amplitude, a linear frequency sweep from 500 Hz to 4 kHz, pure tones at different frequencies, and a final synchronization pulse (see Figure 2):

0 – 1 s: Synchronization pulse: centered around 0.5 s, center frequency is 1 kHz, amplitude 0.9
1 – 2 s: White noise (0.875 s), standard deviation: **10m**, then 0.125 s silence (zeros)
2 – 3 s: White noise (0.875 s), standard deviation: **50m**, then 0.125 s silence (zeros)
3 – 4 s: White noise (0.875 s), standard deviation: **200m**, then 0.125 s silence (zeros)
4—14 s: Linear frequency sweep (10 s), from **100 Hz to 4 kHz**, amplitude 0.9, no silence gap after
14 – 15 s: Pure sine (0.875 s), amplitude 0.9, frequency **250 Hz**, then 0.125 s silence (zeros)
15 – 16 s: Pure sine (0.875 s), amplitude 0.9, frequency **500 Hz**, then 0.125 s silence (zeros)
16 – 17 s: Pure sine (0.875 s), amplitude 0.9, frequency **1 kHz**, then 0.125 s silence (zeros)
17 – 18 s: Pure sine (0.875 s), amplitude 0.9, frequency **2 kHz**, then 0.125 s silence (zeros)
18 – 19 s: Pure sine (0.875 s), amplitude 0.9, frequency **4 kHz**, then 0.125 s silence (zeros)
19 – 20 s: Synchronization pulse: centered around 19.5 s, center frequency is 1 kHz, amplitude 0.9

To broadcast the signal, it is streamed from the host computer to the FPGA where it gets first low-cut filtered to avoid dc components (these would be dangerous for the amplifier and the load speaker). The test signal is resampled from 51.2 kHz to multiples of 20 µs (50 kHz) of the GPS clock for a synchronous recording. This resampling is done by linear interpolation which can cause some slight spectral artefacts like a modulation at 1.2kHz visible in Figure 2, second subpanel.

## Experimental runs

We performed three series of broadcasts from various locations around the fiber with a rowboat. Two runs in the mornings of Jan. 26[th] and 27[th] between 4 am and 6 am in a quiet lake with no ship traffic. The run in the afternoon of Jan. 26th encounters more traffic and noise. For the run in the afternoon we could not deploy the hydrophone recorders. For the first night run, hydrophone recorders A & B did not function properly and are excluded from the datasets. For the second night run, all hydrophone recorders worked and at every location we played the test sequence twice at different depths (2 m, 4 m). For an overview of the second night run, see Figure 1. In summary we have these five runs :

- 2022-01-26--03--Morning
- 2022-01-26--04--Whales
- 2022-01-26--14--Afternoon
- 2022-01-27--2m
- 2022-01-27--4m

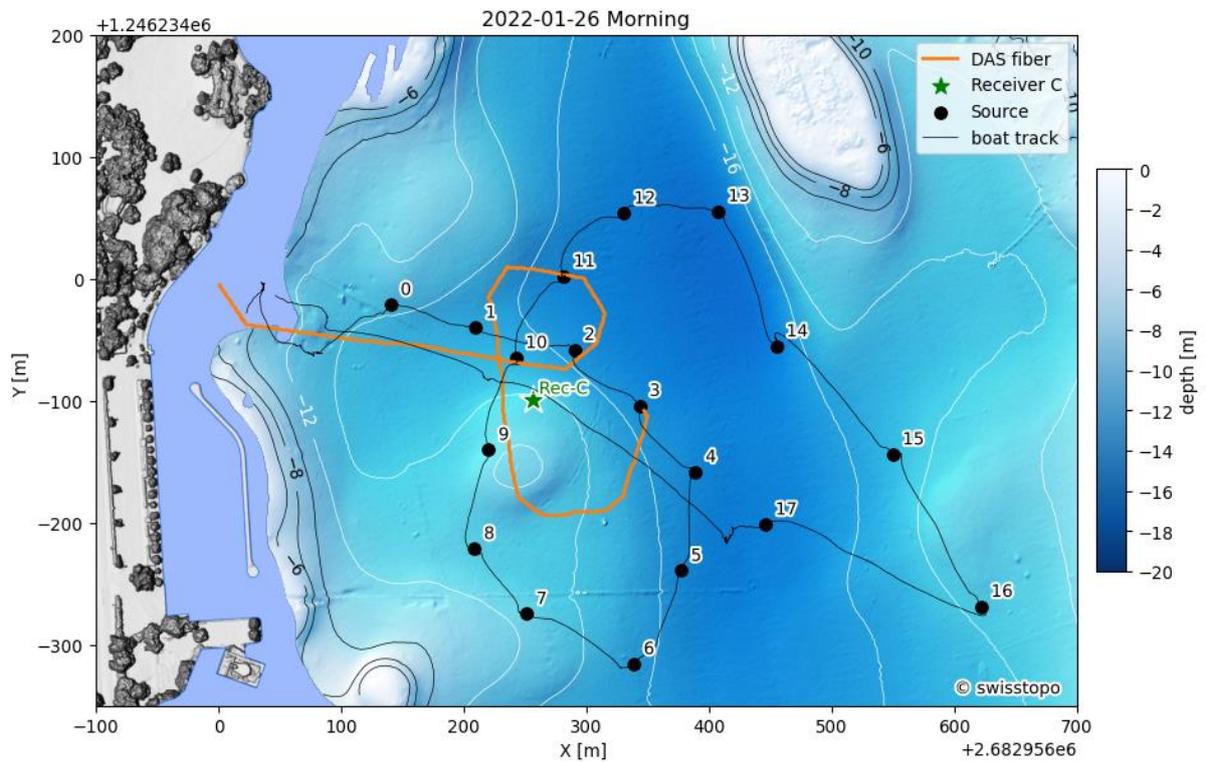

Figure 12. The run of the first morning with 18 source positions. The recorder A & B were not working properly and are excluded from the dataset.

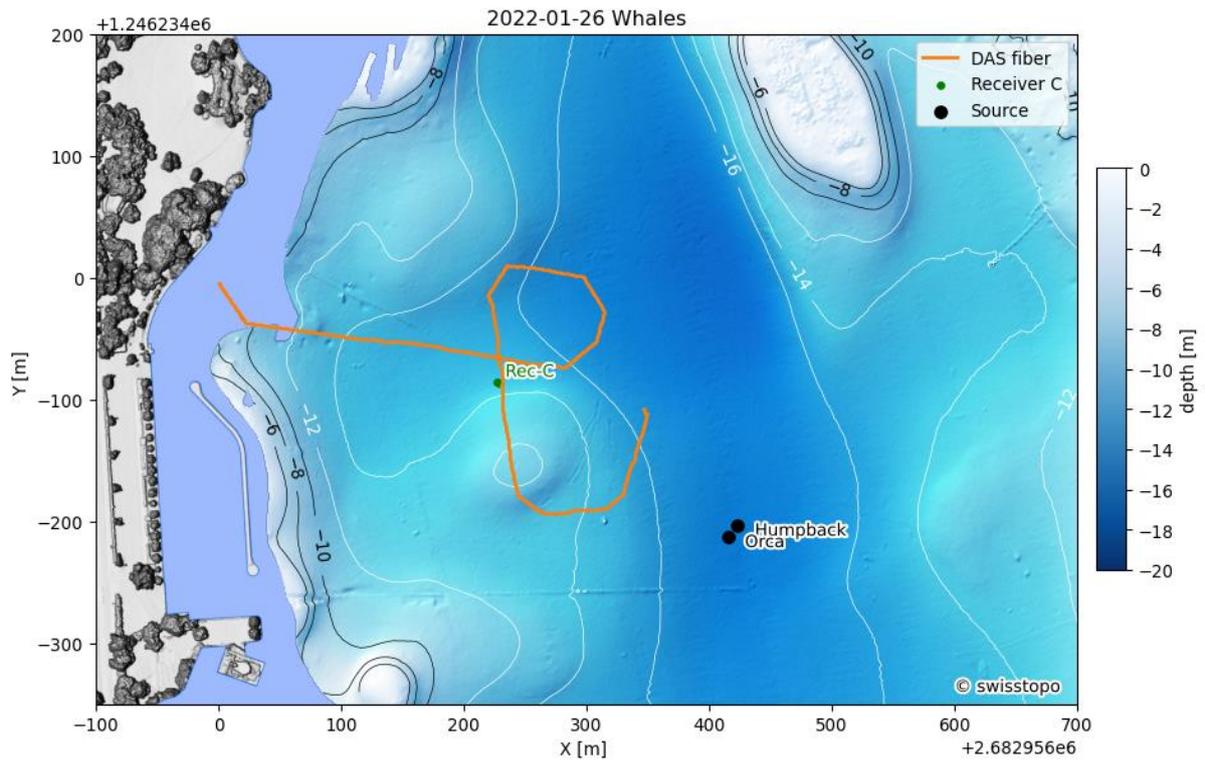

Figure 13. At the end of the run on the first morning we played back recordings from humpback whales and killer whales. With this dataset the algorithms for localization and separation can be tested on natural sounds.

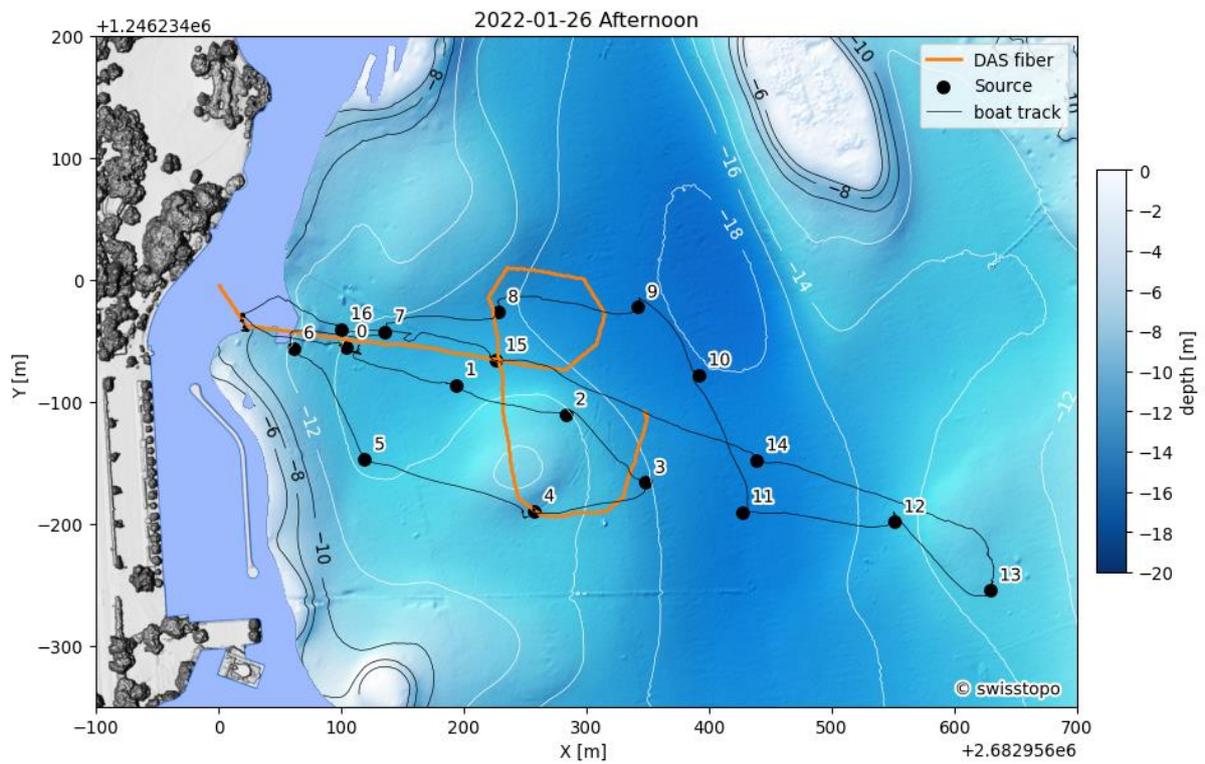

Figure 14. The run on the afternoon of the first day we could not deploy the hydrophone recorder buoys because of ship traffic. This dataset has more noise but is valuable for localization performance evaluations.

## Data management

See the git repository for Jupyter scripts documenting the data management:

https://gitlab.switch.ch/cetacean-communication/expeditions/daslakezh/datapublication.git

The consolidation (or compilation) process had several steps:

- Find the onset of each test signal broadcast and cut out a time segment of 25 s from the DAS recordings and the hydrophone recordings. Combine all signals into one HDF5 file.
- Apply the calibration of the hydrophones and the speaker. This yields a quantitative measurement of sound pressure in the units of Pa.
- Project the geographic information into local cartesian coordinate system. Read in all GPS tracks and provide an estimate of the fiber location. Write all tracks and the bathymetry data of the lake into the dedicated file 'Situation.h5'
- Add all relevant and known metadata to each file.
- Provide example code to read and visualize the data set.

## Filename format

date: year-month-day | time: hour-min-sec | shot nr. | source depth

2022-01-27--04-58-24--27--4m.h5

Figure 15: Composition of the filename. The date and time of the first sample in UTC. The shot number is a consecutive number for each broadcast starting with zero. The last field of characters is for additional information as for example the depth of the source.

## HDF5 file format

To inspect the hdf5 files, use HDF VIEW .

Figure 16: Properties on the file level show general meta data related to the expedition and the data publication itself.

*Figure 17: Properties of the data table "DAS".*

*Figure 18: Object info of the data table "DAS". The data is in 16 bit integers (int16), contains 946 channels and 125'000 samples (25 s * 5 kHz sampling rate).*

*Figure 19: Properties of the data table "Recorder-A". The data is 4 x 125'000 float32: four hydrophones sampled at 50 kHz for 25 s.*

*Figure 20. Properties of the data table "Source". The data is 1 x 125'000 float32: source signal sampled at 50 kHz for 25 s.*

# Acknowledgments

We are particularly grateful to the Zurich Lake Police for (1) helping us with the deployment and recovery of the DAS cable, (2) diving along the entire cable at 4◦C water temperature in order to check the installation, and (3) not arresting us on the spot. This experiment would not have been possible without the support of the 'Seebad Enge' staff who allowed us to convert their office into our data acquisition headquarter. We thank the 'Amt für Abfall, Wasser, Energie und Luft (AWEL)' for the help and quickly issuing a permission.

Jörg Rychen was financially supported by the Swiss National Science Foundation: NCCR Evolving Language Agreement no. 51NF40_180888.

Patrick Paitz was financially supported by ETH Zurich Grant ETH-01-16-2

Krystyna Smolinski was financially supported through the RISE project, funded by the European Union's Horizon 2020 research and innovation program under grant agreement No. 821115.

# References

Barlow, J., Griffiths, E. T., Klinck, H., & Harris, D. V. (2018). Diving behavior of Cuvier's beaked whales inferred from three-dimensional acoustic localization and tracking using a nested array of drifting hydrophone recorders. *The Journal of the Acoustical Society of America*, *144*(4), 2030–2041. https://doi.org/10.1121/1.5055216

Bouffaut, L., Taweesintananon, K., Kriesell, H. J., Robin, A. R., Potter, J. R., Landrø, M., Johansen, E. S., Brenne, J. K., Haukanes, A., Schjelderup, O., & Storvik, F. (2022). Eavesdropping at the speed of light : distributed acoustic sensing of baleen whales in the Arctic. *Frontiers in Marine Science*, *9* (March), 1–13. https://doi.org/10.3389/fmars.2022.901348

Grosso, V. A. Del, & Mader, C. W. (2005). Speed of Sound in Pure Water. *The Journal of the Acoustical Society of America*, *52*(5B), 1442. https://doi.org/10.1121/1.1913258

Miller, B. S., & Dawson, S. (2009). A large-aperture low-cost hydrophone array for tracking whales from small boats. *Journal of the Acoustical Society of America*, *126*(5), 2248–2256. https://doi.org/10.1121/1.3238258

Møhl, B., Wahlberg, M., & Heerfordt, A. (2001). A large-aperture array of nonlinked receivers for acoustic positioning of biological sound sources. *The Journal of the Acoustical Society of America*, *109*(1), 434–437. https://doi.org/10.1121/1.1323462

Rychen, J., Paitz, P., Edme, P., & Fichtner, A. (2023). Test experiments with distributed acoustic sensing and hydrophone arrays for locating underwater sound sources. *Zenodo*. https://doi.org/10.5281/zenodo.7886409

Swisstopo. (2016). *Formulas and constants for the calculation of the Swiss conformal cylindrical projection and for the transformation between coordinate systems*. https://swisstopo.admin.ch

Swisstopo. (2022). *SwissBathy3D*. https://www.swisstopo.admin.ch/de/geodata/height/bathy3d.html